\documentclass{iau} 
\usepackage{graphicx}
\usepackage{natbib}

\def\Msol{\thinspace\hbox{$\hbox{M}_{\odot}$}}

\def\ie{{\it i.e.} }                    
\def\a4{\hsize 17.0cm \vsize 25.cm}

\title[S316 Clusters in bimodal regime] 
{Bimodal regime in young massive clusters leading to subsequent stellar
generations}

\author[W\"unsch et al.]   
{
Richard W\"unsch$^1$,
Jan Palou\v{s}$^1$,
Guillermo Tenorio-Tagle$^2$,
Casiana Mu\~noz-Tu\~n\'on$^3$,
\and
So\v{n}a Ehlerov\'a$^1$
}

\affiliation{$^1$Astronomical Institute of the Czech Academy of Sciences, v.v.i. \\
Bo\v{c}n\'\i\ II 1401, 141 31 Prague, Czech Republic \\
email: {\tt richard@wunsch.cz} \\[\affilskip]
$^2$ Instituto Nacional de Astrof\'\i sica Optica y Electr\'onica \\
AP 51, 72000 Puebla, M\'exico \\[\affilskip]
$^3$ Instituto de Astrof\' \i sica de Canarias \\
38200 La Laguna, Tenerife, Spain
}

\pubyear{2015}
\volume{316}  
\jname{Formation, Evolution, and Survival of Massive Star Clusters}
\editors{C. Charbonnel \& A. Nota, eds.}
\begin{document}

\maketitle

\begin{abstract}
Massive stars in young massive clusters insert tremendous amounts of mass and
energy into their surroundings in the form of stellar winds and supernova
ejecta. Mutual shock-shock collisions lead to formation of hot gas, filling the
volume of the cluster. The pressure of this gas then drives a powerful cluster
wind. However, it has been shown that if the cluster is massive and dense
enough, it can evolve in the so--called bimodal regime, in which the hot gas
inside the cluster becomes thermally unstable and forms dense clumps which are
trapped inside the cluster by its gravity. 
We will review works on the bimodal regime and discuss the implications for the
formation of subsequent stellar generations. The mass accumulates inside the
cluster and as soon as a high enough column density is reached, the interior of
the clumps becomes self-shielded against the ionising radiation of stars and
the clumps collapse and form new stars. The second stellar generation will be
enriched by products of stellar evolution from the first generation, and will
be concentrated near the cluster center.
\end{abstract}

\firstsection 
\section{Introduction}

It has been found by photometric observations that globular clusters contain two
or more populations of stars differing by age and/or the chemical composition
\citep[see e.g.][and references therein]{2000ApJ...534L..83P,
2004ApJ...605L.125B, 2007ApJ...661L..53P, 2015AJ....149...91P}. Furthermore,
spectroscopic observations found anticorrelations between abundances of Na and O
and other pairs of elements \citep{2009A&A...505..139C,2012MNRAS.426.2889M}
suggesting that a certain fraction of stars in globular clusters contain
products of hydrogen burning at high temperatures. Several hypotheses were
suggested to explain the above observations \citep{2007A&A...464.1029D,
2008MNRAS.391..825D, 2013MNRAS.436.2398B, 2009A&A...507L...1D}. Here we propose
a cooling winds scenario in which the second stellar generation is formed out of
fast stellar winds enriched by products of hydrogen burning in massive stars,
that cool down inside the cluster evolving in the so--called bimodal regime.

Young massive clusters with masses $10^{5} - 10^{7}$\,\Msol\ and ages
$<10^{7}$~yr include a high number of massive stars ($\sim 2\times 10^{4}$ per
$10^{6}$\,\Msol\ of the cluster stellar mass assuming the standard IMF)
concentrated in a small volume of several pc in radius \citep[see
e.g.][]{2010ARA&A..48..431P,2010RSPTA.368..867L}. These stars insert through
stellar winds large amounts of gas moving with velocities several thousands
km\,s$^{-1}$ into their surroundings. As the winds collide with each other, their
kinetic energy is thermalised and the gas is heated to temperatures $\sim
10^7$\,K. The high pressure of this hot gas then drives the star cluster wind.

\begin{figure}[t]
\begin{center}
 \includegraphics[width=0.49\textwidth]{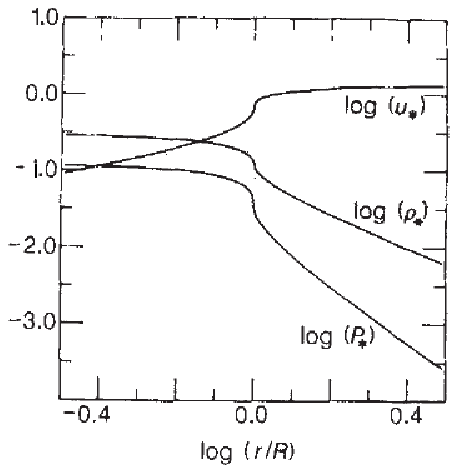} 
 \includegraphics[width=0.49\textwidth]{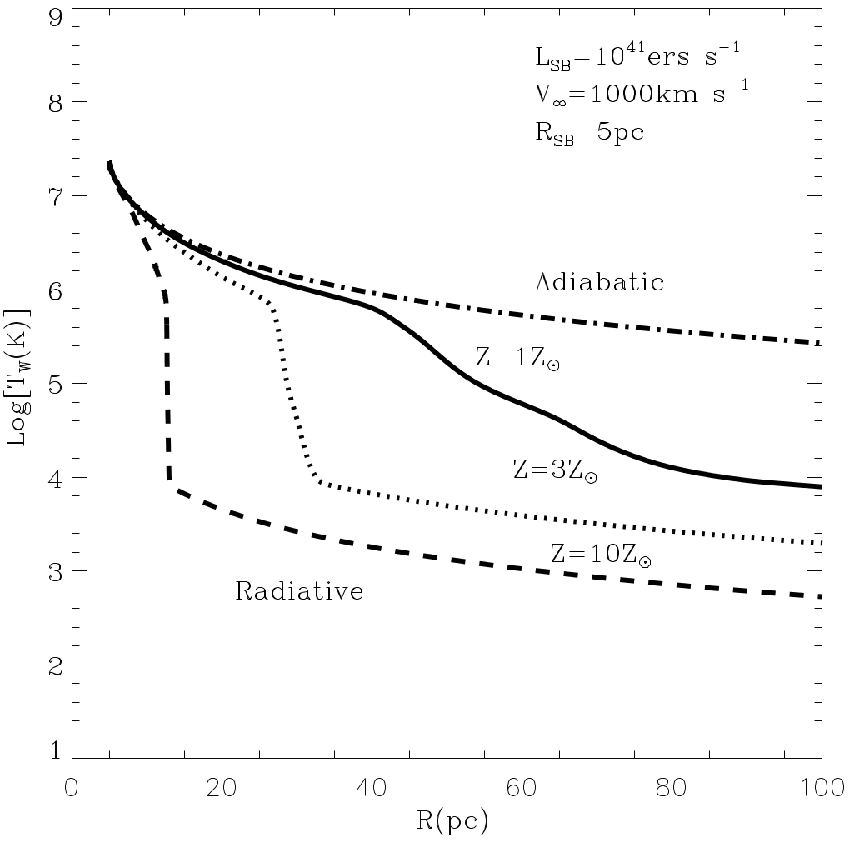}
 \caption{Left: Cluster wind solution by \citet[][Fig.~1, reprinted by 
 permission from Nature Publishing Group]{1985Natur.317...44C}, it shows radial
 profiles of the wind density ($\rho_\star$), pressure ($P_\star$) and velocity
 ($u_\star$) in logarithmic scale. Right: Radial profile of the wind
 temperature, comparison of the adiabatic solution to the radiative solution by
 \citet[][Fig.~1, reprinted by permission from AAS]{2003ApJ...590..791S} for
 three different metallicities.}
 \label{fig:cc85:ss03}
\end{center}
\end{figure}

Winds of young massive star clusters were studied analytically by
\citet{1985Natur.317...44C} who derived a stationary solution of spherically
symmetric hydrodynamic equations. They assumed that sources of mass and energy
are distributed uniformly in a sphere with a given radius and they neglected
radiative cooling of the hot gas. A remarkable property of their solution is
that the wind velocity reaches the sound speed always exactly at the cluster
border. Radial profiles of the basic hydrodynamic quantities given by their
solution are shown on the left panel of Fig.~\ref{fig:cc85:ss03}.

The Chevalier\&Clegg wind solution was later tested and further explored by e.g.
\citet{2000ApJ...536..896C,2001ApJ...559L..33R} and many other authors. An
important field of research is an interaction of the cluster with the ambient
gas, in particular with the remnant of the parent molecular cloud. This subject
is not covered here and we refer e.g. to \citet{2006ApJ...643..186T,
2009ApJ...693.1696H, 2012A&A...546L...5K, 2013A&A...552A.121K,
2013MNRAS.431.1337R, 2015IAUGA..2252184H} and references therein. Another
interesting problem is the interaction of cluster winds with each other. It was
studied by \citet{2007NewAR..51..125T} who show that it can lead to the
formation of the super galactic wind as observed for instance in M82 galaxy.

The influence of the radiative cooling on the star cluster wind was studied by
\citet{2003ApJ...590..791S} who included cooling into the spherically symmetric
wind solution similar to the one by \citet{1985Natur.317...44C}. They show that
the wind rapidly cools at some distance from the cluster, which depend on the
cluster parameters, in particular on its mass and the wind metallicity (see
Fig.~\ref{fig:cc85:ss03}). The radiative solution was compared to X-ray
observations of young clusters by \citep{2004ApJ...610..226S}, for instance it
was shown that it is in good agreement with measured X-ray fluxes of the
nuclear cluster NGC~4303 \citep{2003ApJ...593..127J}.

It has also been found that if the cluster is massive and compact enough, \ie if
its wind mechanical luminosity $L_\mathrm{crit}$ exceeds a certain value, no
stationary wind solution exists \citep{2003ApJ...590..791S}. This can be easily
understood from basic scaling relations. The energy available for driving the
wind is directly proportional to the total energy of stellar winds which is
directly proportional to the cluster mass. On the other hand, energy losses due
to cooling are proportional to the square of the wind density, and therefore to
the square of the cluster mass. As a result, the energy losses will always
dominate if the cluster is massive and compact enough. This led
\citet{2005ApJ...628L..13T} to suggest that the star formation feedback in
massive clusters takes an extreme positive form leading to a high star formation
efficiency.

\section{Bimodal regime and secondary star formation}

\begin{figure}
\begin{center}
 \includegraphics[width=0.43\textwidth]{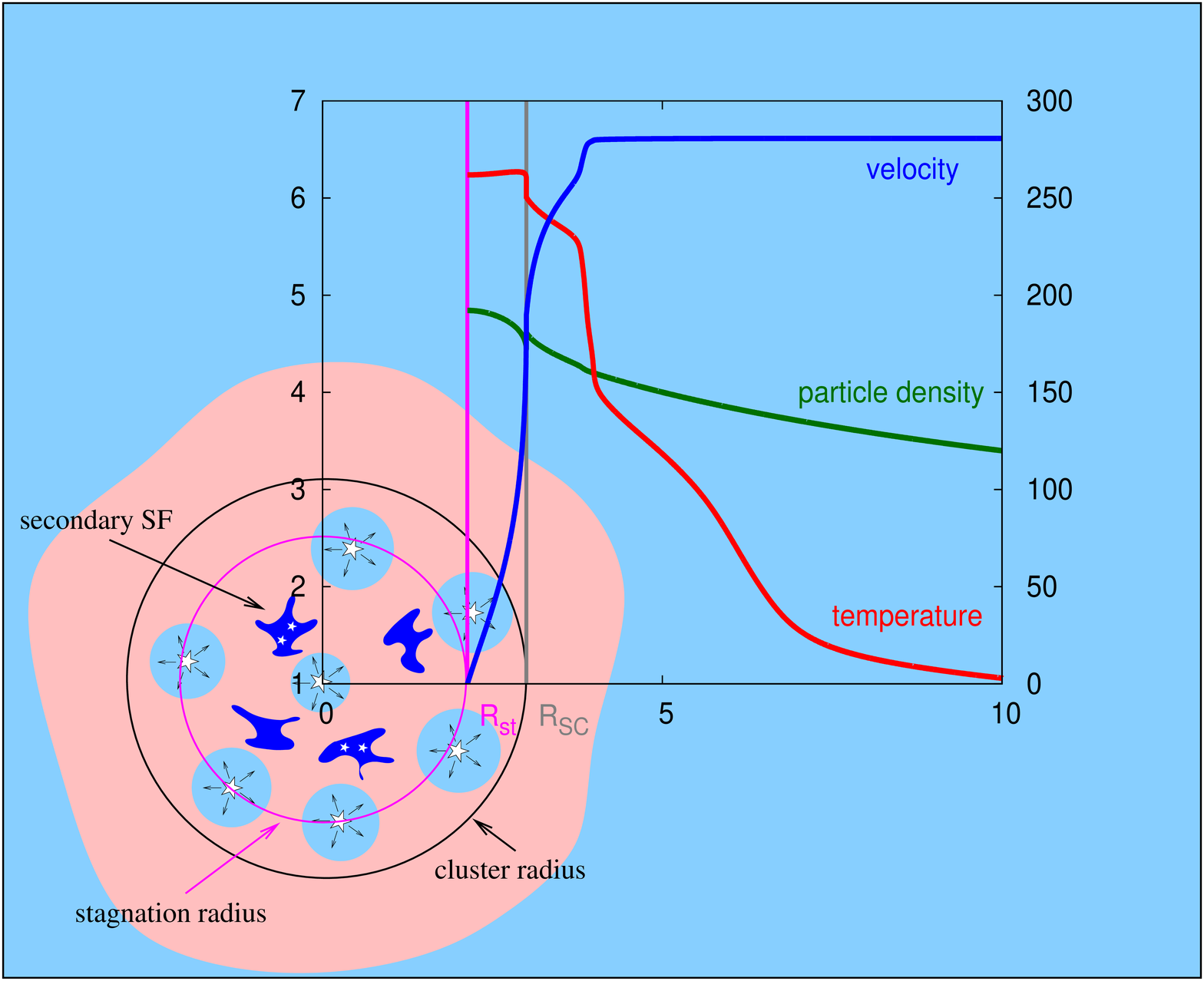}
 \includegraphics[width=0.56\textwidth]{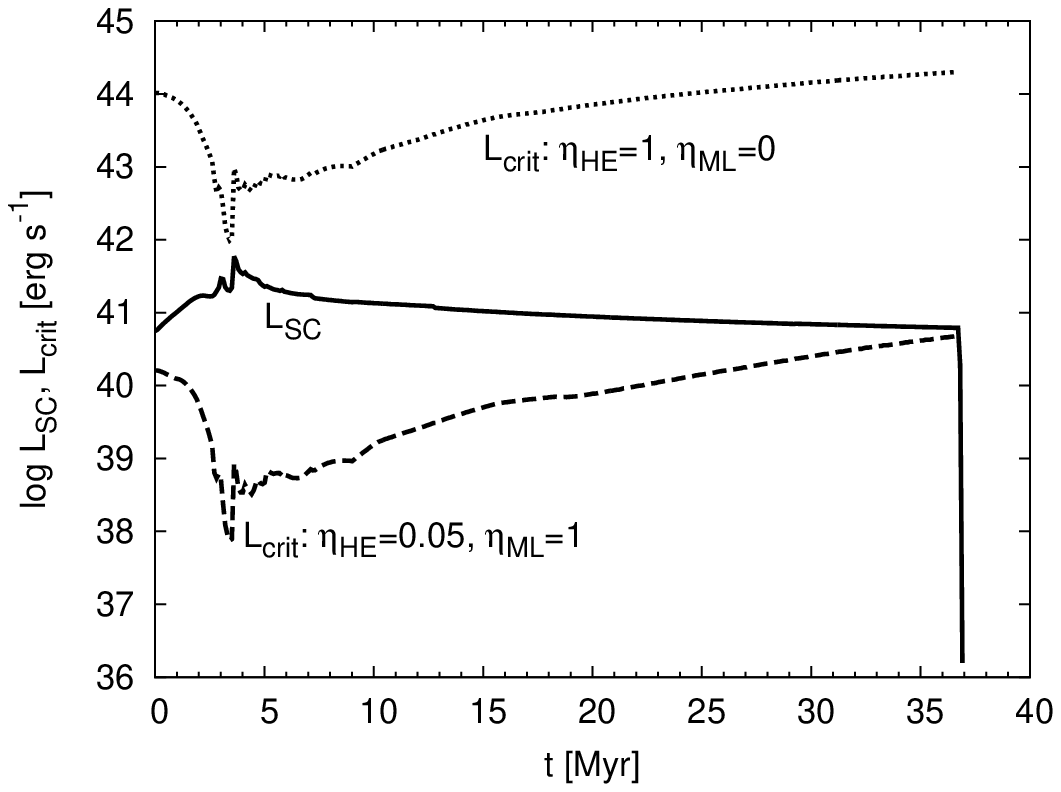}
 \caption{Left: schematic view of the cluster evolving in the bimodal regime.
 Individual stars are surrounded by free wind regions, however, most of the
 cluster volume is filled by the hot gas heated up the wind reverse shock. The
 hot gas is thermally unstable and forms dense warm clumps in some regions inside
 the cluster. The graph shows the wind density, temperature and velocity in the
 outer part of the cluster where the stationary solution exists. Right: time
 evolution of the wind mechanical luminosity (for cluster with mass
 $10^7$\,\Msol, solid) compared to the critical luminosity calculated for heating
 efficiency 1 and no mass loading (dotted) and heating efficiency 5\% and
 mass loading $1$ (dashed).}
 \label{fig:bimod}
\end{center}
\end{figure}

\citet{2007ApJ...658.1196T} studied winds of clusters with mechanical
luminosities higher than $L_\mathrm{crit}$. They found a new, so--called bimodal
regime, for the wind solution, in which the volume of the cluster is divided
into two regions (see Fig.~\ref{fig:bimod} left). In the outer region, the
stationary solution still exists with the wind velocity starting from zero at the
stagnation radius, $R_\mathrm{st}$, and reaching the sound speed at the cluster
border $R_\mathrm{SC}$. In the inner region below $R_\mathrm{st}$, random
parcels of hot gas become thermally unstable, cool down and get compressed by
the ambient hot gas into dense clumps. These dense clumps may stay warm,
maintained at temperature $T\sim 10^4$\,K by the stellar ionising radiation,
however, if their column densities become large enough, they can self-shield
themselves (see Section \ref{sec:radiation}), cool to lower temperatures and
feed the secondary star formation inside the cluster. The existence of the
bimodal regime was confirmed by 2D hydrodynamic simulations by
\citet{2008ApJ...683..683W} where also the estimates of the mass accumulated
inside the cluster were provided. This prediction of the mass accumulation and
the secondary star formation is particularly interesting in context of formation
of globular clusters and their observed multiple stellar populations. 

There is an observational evidence that the temperature of the hot gas inside
young massive clusters is lower than several times $10^7$\,K which correspond to
the thermalisation of all the kinetic energy of individual stellar winds. It led
to the introduction of the so--called heating efficiency, $\eta_\mathrm{HE}$,
denoting a fraction of the stellar wind kinetic energy that is converted into
the hot gas internal energy \citep{2007ApJ...669..952S}. For instance, hydrogen
recombination lines of super star clusters in the Antennae galaxies exhibit
moderately supersonic widths \citep{2007ApJ...668..168G}, larger than individual
velocities of stars, but smaller than typical velocities of stellar winds. They
were interpreted in terms of the bimodal wind solution by
\citet{2010ApJ...708.1621T} who suggest that the line profiles are consistent
with $\eta_\mathrm{HE} \lesssim 0.2$. \citet{2009ApJ...700..931S} found a
similarly small values $\eta_\mathrm{HE} \lesssim 0.1$ by measuring and
analysing sizes of HII regions coinciding with super star clusters in M82
galaxy. On the other hand, \citet{2009ApJ...697.2030S} found relatively high
values of the heating efficiency, $\eta_\mathrm{HE} = 0.3 - 1.0$ by comparing
X-ray observation of M82 with a set of 1D and 2D hydrodynamic models. Another
effect that can decrease the temperature of the hot gas is mass loading of the
wind with the primordial gas. \citet{2009ApJ...697.2030S} estimate the mass
loading factor, $\eta_\mathrm{ML}$, having a moderate values between $1$ and
$2.8$. Additionally, there is evidence coming from observations of the Li
abundance \citep{2007A&A...464.1029D}, that stars in subsequent generations in
globular clusters contain approximately $30$\% of the primordial gas. In
summary, the heating efficiency and the mass loading are important parameters of
the wind solution, however, they are not yet very well constrained.

Evolution of the star cluster wind for the first $40$\,Myr was computed by 
\citet{2011ApJ...740...75W} by combining output from the stellar population
synthesis code Starburst99 \citep{1999ApJS..123....3L} with a semi-analytic code
calculating the spherically symmetric wind solution. Fig.~\ref{fig:bimod}
(right) shows the evolution of the wind mechanical luminosity for a cluster with
mass $M_\mathrm{SC} = 10^7$\,\Msol and radius $R_\mathrm{SC} = 3$\,pc compared
to the evolution of the the critical luminosity $L_\mathrm{crit}$ given for two
combinations of the heating efficiency and the mass loading. It can be seen that
the most conservative and rather unrealistic values $\eta_\mathrm{HE} = 1$ and
$\eta_\mathrm{ML} = 0$ represent a marginal case for which the wind is not
bimodal for all the period shown. On the other hand, a more realistic case with
$\eta_\mathrm{HE} = 0.05$ and $\eta_\mathrm{ML}$ gives a solution in the bimodal
regime (and hence mass accumulation) for the whole period of the existence
of massive stars.

\begin{figure}
\begin{center}
 \includegraphics[width=\textwidth]{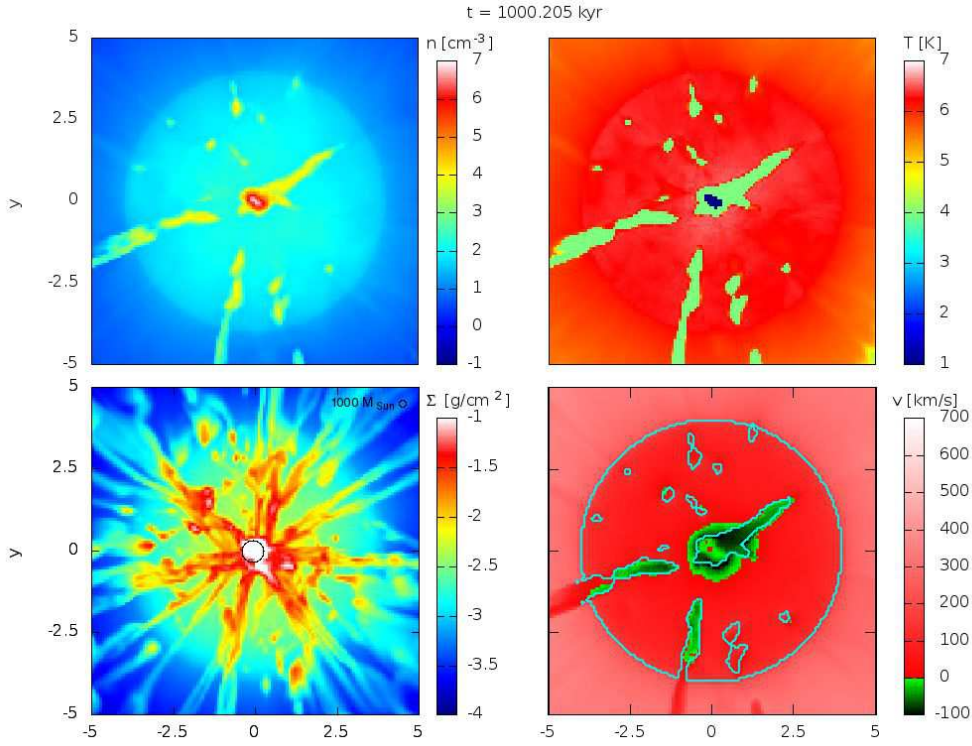}
 \caption{A frame from the 3D radiation-hydrodynamic simulation of a cluster in
 the bimodal regime at age $1$~Myr. Cluster parameters are given in
 Tab.~\ref{tab:clpar}. Individual panels show: logarithm of the gas volume
 density at plane $z=0$ (top left), logarithm of the gas temperature (top right)
 at $z = 0$, logarithm of the gas column density integrated in z-direction
 (bottom left), and the gas radial velocity (bottom right). In the last panel in
 the color version of this figure, the red/green color shows the outward/inward
 velocity and the cyan line marks the points where the flow changes from
 subsonic to supersonic. (see online edition for the colour version)}
 \label{fig:sim}
\end{center}
\end{figure}

\section{Influence of stellar ionising radiation}
\label{sec:radiation}

If the cluster evolves in the bimodal regime, a certain fraction of the gas
reinserted within the cluster by massive star in a form of their winds gets
thermally unstable, cools down and accumulates inside the cluster. However, the
stellar ionising radiation can still maintain the gas warm and ionised if the
column density of dense warm clumps does not exceed the value necessary for
self-shielding. The minimum mass, $m_\mathrm{self}$, of a spherical clump which
is able to self-shield against the ionising radiation was estimated analytically
by \citet{2014ApJ...792..105P}:
\begin{equation}
m_\mathrm{self} = \dot{N}_\mathrm{UV,SC} \frac{\mu m_\mathrm{H}}{\alpha_\star}
\frac{k T_\mathrm{ion}}{P_\mathrm{hot}}
\end{equation}
where $\dot{N}_\mathrm{UV,SC}$ is the total production rate of ionising
photons, $T_\mathrm{ion}$ is the temperature of the warm ionised gas,
$P_\mathrm{hot}$ is the pressure in the hot gas, $\alpha_\star$ is the
recombination coefficient to the second and higher levels, $\mu$ is the
average molecular weight of particles in the hot gas, $m_\mathrm{H}$ is the hydrogen
nuclei mass and $k$ is the Boltzmann constant.

If the gravity of the cluster is taken into account, dense clumps start to sink
into the cluster centre and evolve in more or less steady streams, as shown by
the hydrodynamic simulations (see below). The gas streams collide in the centre
forming a massive central clump. If self-shielding occurs, it may happen
either in both streams and the central clump, or only in the central clump. It
has interesting implications for the second stellar generation, because in the
latter case, it should be very compact, because the stars are formed out of gas
with a very small velocity dispersion. To decide which of the two cases takes
place, time $t_\mathrm{SS}$ needed for self-shielding of the stream can be
estimated and compared to the free fall time $t_\mathrm{ff}$ of the dense clump
into the cluster centre. The self-shielding time is

\begin{equation}
t_\mathrm{SS} = \frac{4\pi^2}{9} N_\mathrm{UV}^2 R_\mathrm{SC}^{-1} 
R_\mathrm{st}^{-2} (1 + \eta_\mathrm{ML})^{-1} 
\dot M_\mathrm{SC}^{-1} \mu m_\mathrm{H} \alpha_*^{-2} \left(\frac{k
T_\mathrm{ion}}{P_\mathrm{hot}}\right)^3 \ .
\end{equation}
In order to test the above model and to explore it in more detail, we run 3D
radiation hydrodynamic simulations of the cluster wind in the bimodal regime. It
is based on the publically available hydrodynamic code Flash
\cite{2000ApJS..131..273F}, regions that are kept warm by the ionising
radiation are determined by our code TreeRay, which works in this setup in a
similar way as the TreeCol algorithm \citep{2012MNRAS.420..745C}. The numerical
model includes radiative cooling, the background gravitational field of the cluster
and self-gravity of the gas. The mass and energy are inserted smoothly in a
spherical volume with a radial distribution given by the Schuster profile
\citep{2013ApJ...772..128P}. More details will be given in the forthcoming
publication (W\"unsch et al., 2015).

\begin{table}
  \begin{center}
  \begin{tabular}{|l|c|c|}
  \hline
  quantity & symbol & value \\
  \hline
  cluster half mass radius & $R_\mathrm{h}$ & $2.67$\,pc \\
  \hline
  stellar mass of the 1st generation & $M_\mathrm{SC}$ & $10^7$\,\Msol \\
  \hline
  mass inserted by stellar winds & $M_\mathrm{sw}$ & $4\times 10^{5}$\,\Msol \\
  \hline
  heating efficiency & $\eta_\mathrm{HE}$ & $0.05$ \\
  \hline
  mass loading & $\eta_\mathrm{ML}$ & $1$ \\
  \hline
  mass loaded & $M_\mathrm{ML}$ & $4\times 10^{5}$\,\Msol \\
  \hline
  mass of the 2nd stellar generation at $3.5$\,Myr & $M_\mathrm{2G}$ & $7\times 10^{5}$\,\Msol \\
  \hline
  mass of the warm gas remaining inside cluster at $3.5$\,Myr & $M_\mathrm{warm}$ & $1.4\times 10^{4}$\,\Msol \\
  \hline
  \end{tabular}
  \end{center}
  \caption{Parameters of the star cluster model for which the presented RHD
  simulation was carried out. The last two lines show simulation results: mass of
  the 2nd stellar generation, \ie mass of all sink particles and the mass
  remaining in the simulation in a form of warm gas.}
  \label{tab:clpar}
\end{table}

Fig.~\ref{fig:sim} shows a snapshot from the simulation with parameters given in
Tab.~\ref{tab:clpar} at time $1$\,Myr. It can be seen that thermally unstable
gas evolves into dense streams that flow into the cluster centre forming there a
massive clump. The streams are fully ionised and stay at $T = 10^4$\,K, however,
the central clump cools to much lower temperatures in its central part. There, the gas
becomes gravitationally unstable and forms sink particles. Due to relatively poor
spatial resolution due to missing proper treatment of thermal and chemical
processes taking place at low temperature, only a few unrealistically massive
sink particles are formed. Therefore, the model is able to predict only the
total mass of the second stellar generation, $M_\mathrm{2G}$, and the
approximate position where the second generation stars are formed. The latter is
interesting in terms of mass loss of the first generation stars, because if the
second stellar generation is highly concentrated, preferential removal of the
first generation stars will be easier independently of the removal mechanism.

\begin{figure}
\begin{center}
 \includegraphics[width=0.48\textwidth]{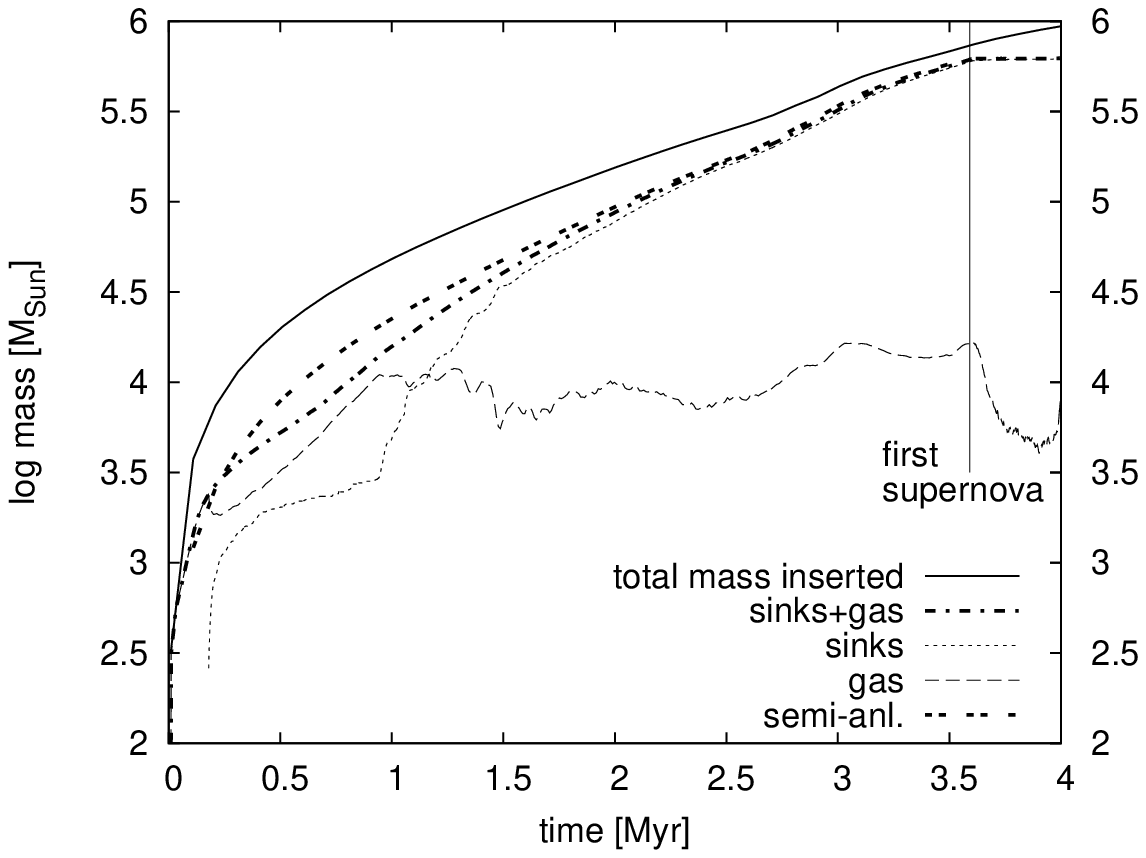}
 \includegraphics[width=0.50\textwidth]{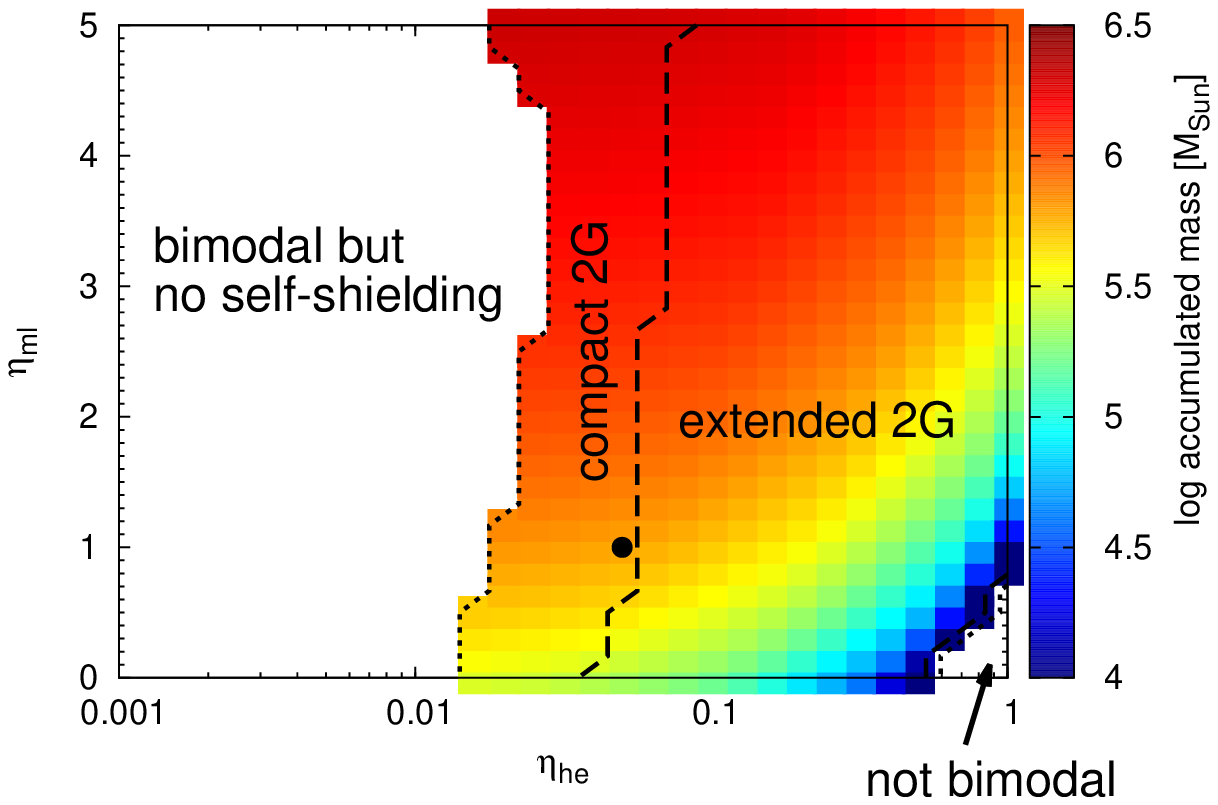}
 \caption{Left: evolution of the total mass inserted into the simulation
 (solid) and its fraction that stays in the simulation (dash-dotted). The latter
 can be split into mass of sink particles (thin dotted) and the mass in the cold/warm
 clumps (thin dashed). Prediction of the accumulated mass by the semi-analytic code
 is shown by the double dashed line. Right: mass of the second stellar generation
 calculated by the semi-analytic code as a function of the heating efficiency,
 $\eta_\mathrm{HE}$, and mass loading, $\eta_\mathrm{ML}$. The second stellar
 generation is predicted to be extended if the thermally unstable gas is able to
 self-shield against ionising radiation before it falls into the centre, and
 compact if it is not. The black circle marks parameters used in the presented
 RHD simulation. (see online edition for the colour version)}
 \label{fig:mevol}
\end{center}
\end{figure}

Evolution of the mass accumulated in the simulation is shown in
Fig.~\ref{fig:mevol} (left) where it is compared to the total mass inserted into
the simulation. The accumulated mass is divided into mass in sink particles and
mass in warm clumps/streams, and it is also compared to the calculation by the
semi-analytic code. It can be seen that the agreement between the complex RHD
simulation and the semi-analytic code is very good. The simulation also includes
a short period after $3.5$\,Myr when supernovae start to explode. As a result,
mass of sink particles stops to grow, \ie star formation is stopped, and the
amount of the warm gas decreases by almost one order of magnitude, \ie most of
the warm gas is removed.

Finally, we explore how properties of the second stellar generation depend on
the heating efficiency $\eta_\mathrm{HE}$ and mass loading $\eta_\mathrm{ML}$.
Fig.~\ref{fig:mevol} (right) shows whether the second generation stars are
formed and if they are, what is their total mass and whether they are formed
only in the central clump (concentrated 2G) or also in the infalling streams
(extended 2G). The mass of the first generation is $10^7$\,\Msol and the
metallicity of the stellar winds is assumed to be solar. The
$\eta_\mathrm{HE}-\eta_\mathrm{ML}$ plane includes four qualitatively different
regions defined mainly by the heating efficiency. If $\eta_\mathrm{HE}$ is very
low, the wind solution is bimodal, however, the accumulated gas is unable to
self-shield against ionising radiation due to its relative low density (given by
a low hot gas pressure). For $0.02 \lesssim \eta_\mathrm{HE} \lesssim 0.07$,
self-shielding is possible only in the central clump and a concentrated second
stellar generation is formed. If $\eta_\mathrm{HE} \gtrsim 0.07$, self-shielding
is possible also in infalling streams and the extended second generation is
formed. In the small region around $\eta_\mathrm{HE} = 1$ and $\eta_\mathrm{ML}
= 0$, the wind solution is not bimodal, and no mass accumulation occurs.

\section{Summary}

We have reviewed properties of the so--called bimodal star cluster wind solution
which are interesting in terms of secondary star formation in young massive
clusters. We have shown that if the cluster is massive enough, the thermal
instability in the hot gas inside the cluster is inevitable. The exact mass
limit depends on the cluster parameters, in particular, on the poorly
constrained heating efficiency $\eta_\mathrm{HE}$ and mass loading
$\eta_\mathrm{ML}$. However, with the most conservative values $\eta_\mathrm{HE}
= 1$ and $\eta_\mathrm{ML} = 0$, the thermal instability and the bimodal regime
should always occur for clusters with masses above $10^7$\,\Msol.

We suggest the cooling winds model as the possible explanation of the multiple
stellar generations in massive star clusters. Numerical simulations show that
clusters evolving in the bimodal regime accumulate inside them mass reinserted
by massive stars in a form of stellar winds. With realistic cluster parameters
(particularly $\eta_\mathrm{HE}$ being not extremely low), the accumulated gas
is able to self-shield against the ionising radiation of massive stars and it
highly probably leads to secondary star formation. The heating efficiency,
$\eta_\mathrm{HE}$, also determines where the accumulated gas becomes
self-shielding and by that whether the second stellar generation is concentrated
(low $\eta_\mathrm{HE}$) or extended throughout the whole cluster (high
$\eta_\mathrm{HE}$). This is interesting in terms of the so--called mass budget
problem, because all models explaining the subsequent stellar populations
observed in globular clusters as stars formed out of winds and outflows of the
first generation stars have to assume that a substantial fraction of the first
generation is removed from the cluster. If the second generation is concentrated
near the cluster centre, the preferential removal of the first generation is
more probable.

\begin{acknowledgments}
We acknowledge support by the Czech Science Foundation grant 209/15/06012 and by
the institutional project RVO:67985815 of the Astronomical Institute, Academy of
Sciences of the Czech Republic.
\end{acknowledgments}

\bibliographystyle{aa}
\bibliography{rw}

\begin{thebibliography}{39}
\expandafter\ifx\csname natexlab\endcsname\relax\def\natexlab#1{#1}\fi

\bibitem[{{Bastian} {et~al.}(2013){Bastian}, {Lamers}, {de Mink}, {Longmore},
  {Goodwin}, \& {Gieles}}]{2013MNRAS.436.2398B}
{Bastian}, N., {Lamers}, H.~J.~G.~L.~M., {de Mink}, S.~E., {et~al.} 2013,
  \mnras, 436, 2398

\bibitem[{{Bedin} {et~al.}(2004){Bedin}, {Piotto}, {Anderson}, {Cassisi},
  {King}, {Momany}, \& {Carraro}}]{2004ApJ...605L.125B}
{Bedin}, L.~R., {Piotto}, G., {Anderson}, J., {et~al.} 2004, \apjl, 605, L125

\bibitem[{{Cant{\'o}} {et~al.}(2000){Cant{\'o}}, {Raga}, \&
  {Rodr{\'{\i}}guez}}]{2000ApJ...536..896C}
{Cant{\'o}}, J., {Raga}, A.~C., \& {Rodr{\'{\i}}guez}, L.~F. 2000, \apj, 536,
  896

\bibitem[{{Carretta} {et~al.}(2009){Carretta}, {Bragaglia}, {Gratton}, \&
  {Lucatello}}]{2009A&A...505..139C}
{Carretta}, E., {Bragaglia}, A., {Gratton}, R., \& {Lucatello}, S. 2009, \aap,
  505, 139

\bibitem[{{Chevalier} \& {Clegg}(1985)}]{1985Natur.317...44C}
{Chevalier}, R.~A. \& {Clegg}, A.~W. 1985, \nat, 317, 44

\bibitem[{{Clark} {et~al.}(2012){Clark}, {Glover}, \&
  {Klessen}}]{2012MNRAS.420..745C}
{Clark}, P.~C., {Glover}, S.~C.~O., \& {Klessen}, R.~S. 2012, \mnras, 420, 745

\bibitem[{{de Mink} {et~al.}(2009){de Mink}, {Pols}, {Langer}, \&
  {Izzard}}]{2009A&A...507L...1D}
{de Mink}, S.~E., {Pols}, O.~R., {Langer}, N., \& {Izzard}, R.~G. 2009, \aap,
  507, L1

\bibitem[{{Decressin} {et~al.}(2007){Decressin}, {Meynet}, {Charbonnel},
  {Prantzos}, \& {Ekstr{\"o}m}}]{2007A&A...464.1029D}
{Decressin}, T., {Meynet}, G., {Charbonnel}, C., {Prantzos}, N., \&
  {Ekstr{\"o}m}, S. 2007, \aap, 464, 1029

\bibitem[{{D'Ercole} {et~al.}(2008){D'Ercole}, {Vesperini}, {D'Antona},
  {McMillan}, \& {Recchi}}]{2008MNRAS.391..825D}
{D'Ercole}, A., {Vesperini}, E., {D'Antona}, F., {McMillan}, S.~L.~W., \&
  {Recchi}, S. 2008, \mnras, 391, 825

\bibitem[{{Fryxell} {et~al.}(2000){Fryxell}, {Olson}, {Ricker}, {Timmes},
  {Zingale}, {Lamb}, {MacNeice}, {Rosner}, {Truran}, \&
  {Tufo}}]{2000ApJS..131..273F}
{Fryxell}, B., {Olson}, K., {Ricker}, P., {et~al.} 2000, \apjs, 131, 273

\bibitem[{{Gilbert} \& {Graham}(2007)}]{2007ApJ...668..168G}
{Gilbert}, A.~M. \& {Graham}, J.~R. 2007, \apj, 668, 168

\bibitem[{{Harper-Clark} \& {Murray}(2009)}]{2009ApJ...693.1696H}
{Harper-Clark}, E. \& {Murray}, N. 2009, \apj, 693, 1696

\bibitem[{{Herrera} \& {Boulanger}(2015)}]{2015IAUGA..2252184H}
{Herrera}, C.~N. \& {Boulanger}, F. 2015, IAU General Assembly, 22, 52184

\bibitem[{{Jim{\'e}nez-Bail{\'o}n} {et~al.}(2003){Jim{\'e}nez-Bail{\'o}n},
  {Santos-Lle{\'o}}, {Mas-Hesse}, {Guainazzi}, {Colina}, {Cervi{\~n}o}, \&
  {Gonz{\'a}lez Delgado}}]{2003ApJ...593..127J}
{Jim{\'e}nez-Bail{\'o}n}, E., {Santos-Lle{\'o}}, M., {Mas-Hesse}, J.~M.,
  {et~al.} 2003, \apj, 593, 127

\bibitem[{{Krause} {et~al.}(2013){Krause}, {Charbonnel}, {Decressin}, {Meynet},
  \& {Prantzos}}]{2013A&A...552A.121K}
{Krause}, M., {Charbonnel}, C., {Decressin}, T., {Meynet}, G., \& {Prantzos},
  N. 2013, \aap, 552, A121

\bibitem[{{Krause} {et~al.}(2012){Krause}, {Charbonnel}, {Decressin}, {Meynet},
  {Prantzos}, \& {Diehl}}]{2012A&A...546L...5K}
{Krause}, M., {Charbonnel}, C., {Decressin}, T., {et~al.} 2012, \aap, 546, L5

\bibitem[{{Larsen}(2010)}]{2010RSPTA.368..867L}
{Larsen}, S.~S. 2010, Royal Society of London Philosophical Transactions Series
  A, 368, 867

\bibitem[{{Leitherer} {et~al.}(1999){Leitherer}, {Schaerer}, {Goldader},
  {Gonz{\'a}lez Delgado}, {Robert}, {Kune}, {de Mello}, {Devost}, \&
  {Heckman}}]{1999ApJS..123....3L}
{Leitherer}, C., {Schaerer}, D., {Goldader}, J.~D., {et~al.} 1999, \apjs, 123,
  3

\bibitem[{{Mucciarelli} {et~al.}(2012){Mucciarelli}, {Bellazzini}, {Ibata},
  {Merle}, {Chapman}, {Dalessandro}, \& {Sollima}}]{2012MNRAS.426.2889M}
{Mucciarelli}, A., {Bellazzini}, M., {Ibata}, R., {et~al.} 2012, \mnras, 426,
  2889

\bibitem[{{Palou{\v s}} {et~al.}(2013){Palou{\v s}}, {W{\"u}nsch},
  {Mart{\'{\i}}nez-Gonz{\'a}lez}, {Hueyotl-Zahuantitla}, {Silich}, \&
  {Tenorio-Tagle}}]{2013ApJ...772..128P}
{Palou{\v s}}, J., {W{\"u}nsch}, R., {Mart{\'{\i}}nez-Gonz{\'a}lez}, S.,
  {et~al.} 2013, \apj, 772, 128

\bibitem[{{Palou{\v s}} {et~al.}(2014){Palou{\v s}}, {W{\"u}nsch}, \&
  {Tenorio-Tagle}}]{2014ApJ...792..105P}
{Palou{\v s}}, J., {W{\"u}nsch}, R., \& {Tenorio-Tagle}, G. 2014, \apj, 792,
  105

\bibitem[{{Pancino} {et~al.}(2000){Pancino}, {Ferraro}, {Bellazzini}, {Piotto},
  \& {Zoccali}}]{2000ApJ...534L..83P}
{Pancino}, E., {Ferraro}, F.~R., {Bellazzini}, M., {Piotto}, G., \& {Zoccali},
  M. 2000, \apjl, 534, L83

\bibitem[{{Piotto} {et~al.}(2007){Piotto}, {Bedin}, {Anderson}, {King},
  {Cassisi}, {Milone}, {Villanova}, {Pietrinferni}, \&
  {Renzini}}]{2007ApJ...661L..53P}
{Piotto}, G., {Bedin}, L.~R., {Anderson}, J., {et~al.} 2007, \apjl, 661, L53

\bibitem[{{Piotto} {et~al.}(2015){Piotto}, {Milone}, {Bedin}, {Anderson},
  {King}, {Marino}, {Nardiello}, {Aparicio}, {Barbuy}, {Bellini}, {Brown},
  {Cassisi}, {Cool}, {Cunial}, {Dalessandro}, {D'Antona}, {Ferraro}, {Hidalgo},
  {Lanzoni}, {Monelli}, {Ortolani}, {Renzini}, {Salaris}, {Sarajedini}, {van
  der Marel}, {Vesperini}, \& {Zoccali}}]{2015AJ....149...91P}
{Piotto}, G., {Milone}, A.~P., {Bedin}, L.~R., {et~al.} 2015, \aj, 149, 91

\bibitem[{{Portegies Zwart} {et~al.}(2010){Portegies Zwart}, {McMillan}, \&
  {Gieles}}]{2010ARA&A..48..431P}
{Portegies Zwart}, S.~F., {McMillan}, S.~L.~W., \& {Gieles}, M. 2010, \araa,
  48, 431

\bibitem[{{Raga} {et~al.}(2001){Raga}, {Vel{\'a}zquez}, {Cant{\'o}},
  {Masciadri}, \& {Rodr{\'{\i}}guez}}]{2001ApJ...559L..33R}
{Raga}, A.~C., {Vel{\'a}zquez}, P.~F., {Cant{\'o}}, J., {Masciadri}, E., \&
  {Rodr{\'{\i}}guez}, L.~F. 2001, \apjl, 559, L33

\bibitem[{{Rogers} \& {Pittard}(2013)}]{2013MNRAS.431.1337R}
{Rogers}, H. \& {Pittard}, J.~M. 2013, \mnras, 431, 1337

\bibitem[{{Silich} {et~al.}(2003){Silich}, {Tenorio-Tagle}, \&
  {Mu{\~n}oz-Tu{\~n}{\'o}n}}]{2003ApJ...590..791S}
{Silich}, S., {Tenorio-Tagle}, G., \& {Mu{\~n}oz-Tu{\~n}{\'o}n}, C. 2003, \apj,
  590, 791

\bibitem[{{Silich} {et~al.}(2007){Silich}, {Tenorio-Tagle}, \&
  {Mu{\~n}oz-Tu{\~n}{\'o}n}}]{2007ApJ...669..952S}
{Silich}, S., {Tenorio-Tagle}, G., \& {Mu{\~n}oz-Tu{\~n}{\'o}n}, C. 2007, \apj,
  669, 952

\bibitem[{{Silich} {et~al.}(2004){Silich}, {Tenorio-Tagle}, \&
  {Rodr{\'{\i}}guez-Gonz{\'a}lez}}]{2004ApJ...610..226S}
{Silich}, S., {Tenorio-Tagle}, G., \& {Rodr{\'{\i}}guez-Gonz{\'a}lez}, A. 2004,
  \apj, 610, 226

\bibitem[{{Silich} {et~al.}(2009){Silich}, {Tenorio-Tagle}, {Torres-Campos},
  {Mu{\~n}oz-Tu{\~n}{\'o}n}, {Monreal-Ibero}, \& {Melo}}]{2009ApJ...700..931S}
{Silich}, S., {Tenorio-Tagle}, G., {Torres-Campos}, A., {et~al.} 2009, \apj,
  700, 931

\bibitem[{{Strickland} \& {Heckman}(2009)}]{2009ApJ...697.2030S}
{Strickland}, D.~K. \& {Heckman}, T.~M. 2009, \apj, 697, 2030

\bibitem[{{Tenorio-Tagle} {et~al.}(2006){Tenorio-Tagle},
  {Mu{\~n}oz-Tu{\~n}{\'o}n}, {P{\'e}rez}, {Silich}, \&
  {Telles}}]{2006ApJ...643..186T}
{Tenorio-Tagle}, G., {Mu{\~n}oz-Tu{\~n}{\'o}n}, C., {P{\'e}rez}, E., {Silich},
  S., \& {Telles}, E. 2006, \apj, 643, 186

\bibitem[{{Tenorio-Tagle} {et~al.}(2007{\natexlab{a}}){Tenorio-Tagle},
  {Silich}, \& {Mu{\~n}oz-Tu{\~n}{\'o}n}}]{2007NewAR..51..125T}
{Tenorio-Tagle}, G., {Silich}, S., \& {Mu{\~n}oz-Tu{\~n}{\'o}n}, C.
  2007{\natexlab{a}}, \nar, 51, 125

\bibitem[{{Tenorio-Tagle} {et~al.}(2005){Tenorio-Tagle}, {Silich},
  {Rodr{\'{\i}}guez-Gonz{\'a}lez}, \&
  {Mu{\~n}oz-Tu{\~n}{\'o}n}}]{2005ApJ...628L..13T}
{Tenorio-Tagle}, G., {Silich}, S., {Rodr{\'{\i}}guez-Gonz{\'a}lez}, A., \&
  {Mu{\~n}oz-Tu{\~n}{\'o}n}, C. 2005, \apjl, 628, L13

\bibitem[{{Tenorio-Tagle} {et~al.}(2010){Tenorio-Tagle}, {W{\"u}nsch},
  {Silich}, {Mu{\~n}oz-Tu{\~n}{\'o}n}, \& {Palou{\v s}}}]{2010ApJ...708.1621T}
{Tenorio-Tagle}, G., {W{\"u}nsch}, R., {Silich}, S., {Mu{\~n}oz-Tu{\~n}{\'o}n},
  C., \& {Palou{\v s}}, J. 2010, \apj, 708, 1621

\bibitem[{{Tenorio-Tagle} {et~al.}(2007{\natexlab{b}}){Tenorio-Tagle},
  {W{\"u}nsch}, {Silich}, \& {Palou{\v s}}}]{2007ApJ...658.1196T}
{Tenorio-Tagle}, G., {W{\"u}nsch}, R., {Silich}, S., \& {Palou{\v s}}, J.
  2007{\natexlab{b}}, \apj, 658, 1196

\bibitem[{{W{\"u}nsch} {et~al.}(2011){W{\"u}nsch}, {Silich}, {Palou{\v s}},
  {Tenorio-Tagle}, \& {Mu{\~n}oz-Tu{\~n}{\'o}n}}]{2011ApJ...740...75W}
{W{\"u}nsch}, R., {Silich}, S., {Palou{\v s}}, J., {Tenorio-Tagle}, G., \&
  {Mu{\~n}oz-Tu{\~n}{\'o}n}, C. 2011, \apj, 740, 75

\bibitem[{{W{\"u}nsch} {et~al.}(2008){W{\"u}nsch}, {Tenorio-Tagle}, {Palou{\v
  s}}, \& {Silich}}]{2008ApJ...683..683W}
{W{\"u}nsch}, R., {Tenorio-Tagle}, G., {Palou{\v s}}, J., \& {Silich}, S. 2008,
  \apj, 683, 683

\end{thebibliography}


\end{document}